\documentclass[10pt,twocolumn]{article}

\usepackage[utf8]{inputenc}
\usepackage[T1]{fontenc}

\usepackage{graphicx}
\usepackage{pdfpages}

\usepackage{ulem}
\usepackage{lipsum}
\usepackage{xcolor}

\usepackage{authblk}
\usepackage[margin=2cm]{geometry}
\usepackage{mathtools, cuted}

\usepackage{bm}
\usepackage{amsfonts}
\usepackage{amsmath}
\usepackage{amssymb}
\usepackage{mathrsfs} 
\usepackage{siunitx}

\usepackage{soul}

\title{Straight to zigzag transition of foam pseudo Plateau borders on textured surfaces}
\author[1]{Alexis Commereuc}
\author[1]{Sandrine Mariot}
\author[1]{Emmanuelle Rio}
\author[1]{François Boulogne}
\affil[1]{Université Paris-Saclay, CNRS, Laboratoire de Physique des Solides, 91405, Orsay, France.}

\date{\today}
\begin{document}

\twocolumn[
    \begin{@twocolumnfalse}
        \maketitle
        \begin{abstract}
            The structure of liquid foams follows simple geometric rules formulated by Plateau 150 years ago.
            By placing such foam on a microtextured hydrophilic surface, we show that the bubble footprint exhibits a morphological transition.
            This transition concerns the liquid channels, also called pseudo Plateau borders, which are straight between vertices on a smooth surface.
            We demonstrate experimentally that for a sufficiently large roughness size compared to the width of the liquid channels, the footprint adopts a zigzag shape.
            This transition is associated with the absence of a wetting film between the pillars caused by capillary suction of the foam, observed by confocal microscopy.
            We rationalize the number of zigzag segments by a geometric distribution describing the observations made with the footprint perimeter and the mesh size of the asperities.
        \end{abstract}
    \end{@twocolumnfalse}
]

%
%
\section{Introduction}

The structure of dry liquid foams adheres to rules originally formulated by Plateau \cite{Plateau1873} and subsequently formally proven by Taylor \cite{Taylor1976}.
These rules can be verified under the assumption that the material is at equilibrium and that the system's energy is proportional to the surface area of the liquid films.
Three fundamental laws govern the behavior of such foams \cite{Weaire1999,Cantat2013b}.
Firstly, the constant curvature of a soap film arises as a result of the difference in Laplace pressures between the two adjacent bubbles it separates.
Secondly, soap films come together in groups of three, forming a liquid channel called Plateau border.
Lastly, four Plateau borders converge to a single point referred to as a vertex.

Experimental observations in the bulk of dry foams confirmed these laws and evidenced that bubbles have between 11 and 17 faces \cite{Matzke1946}.
Recently, Plateau's laws have been altered by the addition of elastic ribbons in the foam~\cite{Jouanlanne2022}.
Alternatively, departure from Plateau's laws is observed for foams made with an emulsion bringing elasticity to the continuous phase~\cite{Guidolin2023}.

When a foam is placed in contact with a hydrophilic surface, the foam structure must comply with this boundary condition.
The foam films establish connections with the surface through straight liquid channels formed by two menisci.
These channels are called pseudo Plateau borders (PPBs).
Therefore, bubble footprints are polyhedral with straight edges between vertices.
Surface bubbles predominantly exhibit a hexagonal shape, although bubbles with five and seven segments are also frequently observed in 3D-foams \cite{Matzke1946} as well as in 2D-foams \cite{Glazier1987,Roth2013}.

Considering that the structure of foam on a surface is influenced by its wetting properties, it is reasonable to anticipate that the surface features can impact the bubble footprints.
Among the various methods available to manipulate surface properties, the introduction of textures has proven to have a significant effect on drop spreading behavior \cite{Cubaud2001, Bico2001}.
While drops on a smooth, homogeneous surface exhibit a circular footprint, surfaces decorated with regularly spaced pillars cause distortion of the contact line along the predominant orientations of the textures, leading to polyhedral shapes \cite{Courbin2007, Kim2011b, Gauthier2014, Kim2016}.

Extensive research has been conducted on the dynamics of drops and contact lines in relation to such surface features \cite{DeGennes1985,Raphael1989,Quere2008a,Gauthier2013,Mannetje2014}.
However, understanding the impact of surface textures on a more complex entity like foams necessitates further investigation \cite{Johnson2022}.
In this Letter, we aim to explore experimentally the role of surface textures on the static footprint of a monodisperse foam.

%
%
\section{Experimental methods}

We prepare a soap solution by diluting a commercial surfactant (Fairy with a concentration in surfactant: 5–15 \%) at a concentration of 10 wt.\% in pure water.
The surface tension is $\gamma = 24.5 \pm 0.1$ mN/m.
Fluorescein is added to the soap solution for visualizations by fluorescence microscopy at a concentration of 1~g/kg.

To generate a foam, the solution is poured in a container with a glass window for visualization on a vertical side and needles pointing upward at the bottom.
The needles, either 22, 27 or 32 Ga, are used to inject air with a pressure controller (OF1, Elveflow, France).
We obtain a monodisperse foam in contact with the soap solution and we stop the bubbling \cite{Marchand2020}.
We selected four values of the bubble radius $R$, which are $[660, 800,1000, 1400]$~$\mathrm{\mu m}$.


The liquid fraction profile is given by $\varphi(z) =  \hat\varphi \left( z/\ell_{\rm c} + \left( \varphi_{\rm c} / \hat\varphi\right)^{-1/2} \right)^{-2}$ \cite{Boulogne2023}, where $\varphi_{\rm c} = 0.26$ is the fraction of gaps in a close-packing of hard spheres and $\hat\varphi = \ell_{\rm c}^2 / R^2\delta^2$ with $\ell_{\rm c} = \sqrt{\gamma / \rho g}$ the capillary length and $\delta = 1.73$ a geometric constant.
In our experiments, we explored liquid fractions between 0.013~\% and 2.4~\%.

Textured surfaces are produced by molding polydimethylsiloxane (PDMS) on a textured surface produced by optical lithography with SU8 resist.
The pattern consists of cubic pillars of edge length $a$ arranged on a square lattice with a spacing equal to the pillar size $a$.
We used seven surfaces for $a\in[0, 30, 60, 100, 130, 160, 200]~\mu$m that are made hydrophilic with a preliminary plasma treatment.
A textured surface is then placed inside the container against the glass window.
Visualizations are made through the textured surface with a custom horizontal fluorescence microscope composed of a long working distance objective (Mitutoyo M Plan Apo x2), a tube lens (Edmund, MT-1), a FITC filter cube (Edmund optics), a LED light source (M470L5, Thorlabs), and a camera (ORCA-Flash4.0 V3, Hamamatsu).

%
%
\section{Results and discussion}

\begin{figure}[h!]
    \centering
    \includegraphics[width=\linewidth]{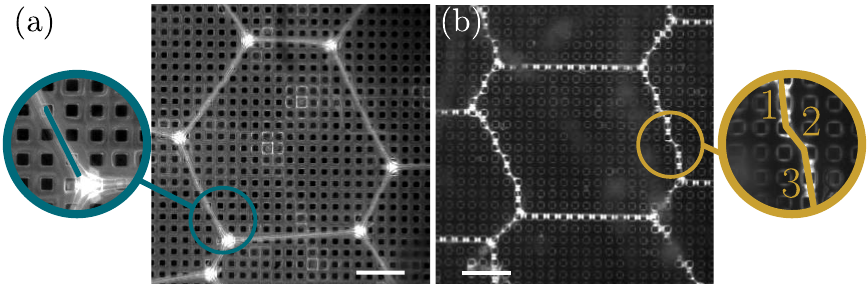}
    \includegraphics[width=\linewidth]{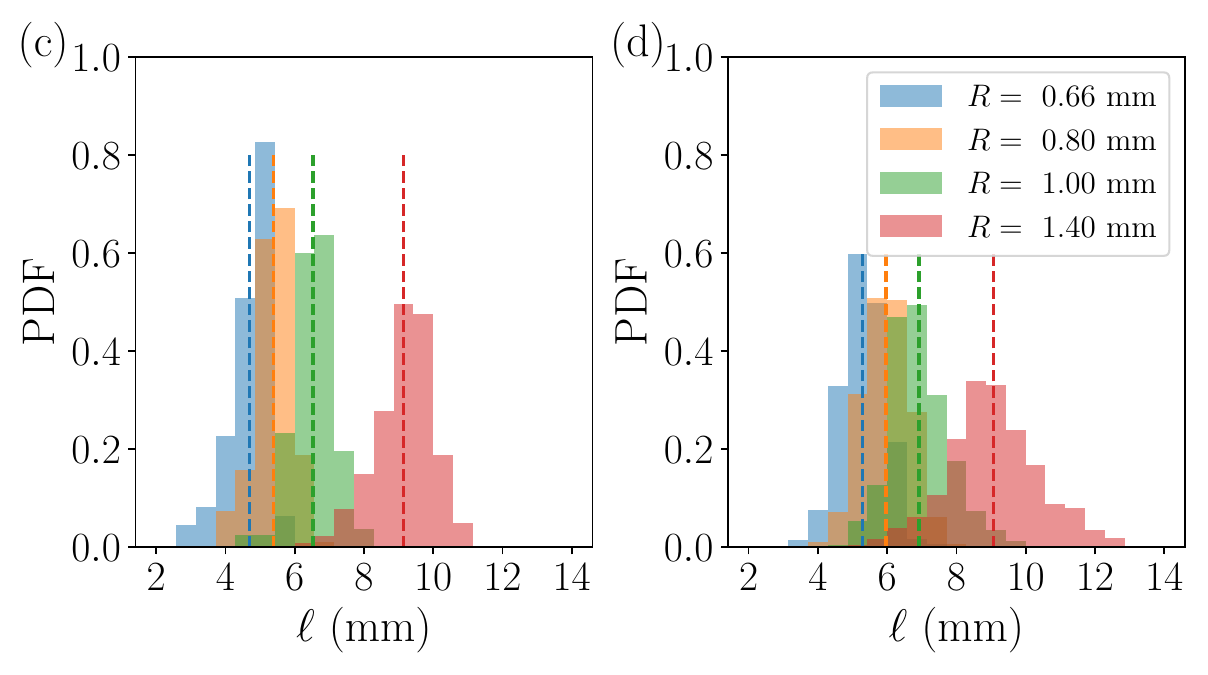}
    \includegraphics[width=\linewidth]{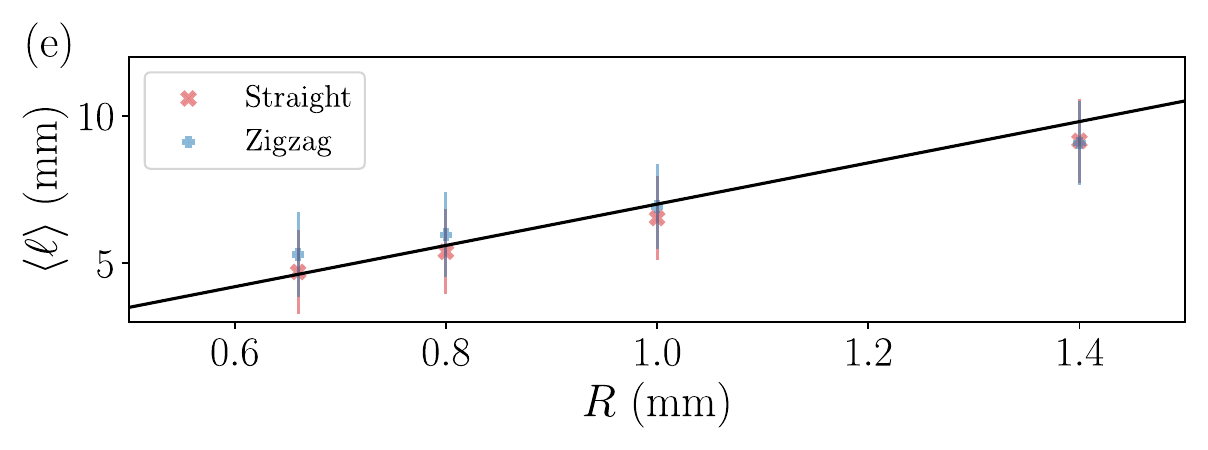}
    \caption{Example of footprint zigzag transition of a monodisperse foam $R= 1000$~$\mathrm{\mu m}$ on a textured surface $a = 60$~$\mathrm{\mu m}$ (High resolution with segment labeling is provided in SI).
    The image (a) is for a liquid fraction $\varphi = 0.28$~\% (\textit{i.e.} $a/r_{\rm pb}=0.65$) where straight edge footprints are observed.
    In (b), for $\varphi = 0.03$~\% (\textit{i.e.} $a/r_{\rm pb}=2.0$), edges have a zigzag morphology.
    The yellow line shows the recorded path of the PPBs constituted of segments.
    Scale bars in white represents 0.6 mm.
    Histograms are the PDF of the footprint perimeters $\ell$ in the (c) straight and (d)  zigzag regimes, respectively.
    The vertical dashed lines are the mean values, which are represented in (e) as a function of $R$.
    The black line of equation $\langle \ell \rangle = 7 R$ is a guide for the eye.
}
    \label{fig:visualization}
\end{figure}

Depending on the experimental parameters $(R, a, \varphi)$, we observe  by fluorescent microscopy different shapes of the bubble footprints on the surfaces.
A first type is presented in figure~\ref{fig:visualization}(a) where the polyhedral footprints are composed of vertices, rendered as bright spots, connected by straight liquid channels, namely the PPBs.
Although the picture in figure~\ref{fig:visualization}(a) is taken on a rough surface, the footprint morphology is identical to contacts on smooth surfaces.
Figure~\ref{fig:visualization}(b) is obtained for a lower liquid fraction.
In this case, the number of vertices remains unchanged but the PPBs are distorted to comply with the surface textures, forming a zigzag morphology.
Our objective is to rationalize the transition from the footprint following the Plateau rules to the zigzag PPBs.
In addition, we aim to characterize the distortion induced by the square lattice as a function of foam and surface properties.

From the photographs, we measured for individual footprints the set $\{\ell_i\}$ composed of the lengths  of each segment illustrated in figure~\ref{fig:visualization}(a, b).
We denote  $n$ the cardinality of the set  $\{\ell_i\}$, which will be referred to the number of segments. Also, the bubble footprint perimeter is $\ell=\sum \ell_i$.
By varying the experimental configurations $(R, a, \varphi)$, we obtained a dataset composed of about 4~000 footprints with typically about 25 footprints per configuration.


The footprint perimeter distributions for the different bubble radii are represented in  figure~\ref{fig:visualization}(c,d) for the straight and zigzag morphologies respectively.
Distributions are similar with nearly identical mean and standard deviation values as shown in figure~\ref{fig:visualization}(e), which shows as well the proportionality between the mean perimeter and the bubble radius.
Thus, the distortion of the edges in the zigzag regime has a weak effect on the footprint perimeter, which is attributed to the separation of lengthscales $\{a, r_{\rm pb}\} \ll \{\ell, R\}$.


\begin{figure}[ht]
    \centering

    \includegraphics[width=\linewidth]{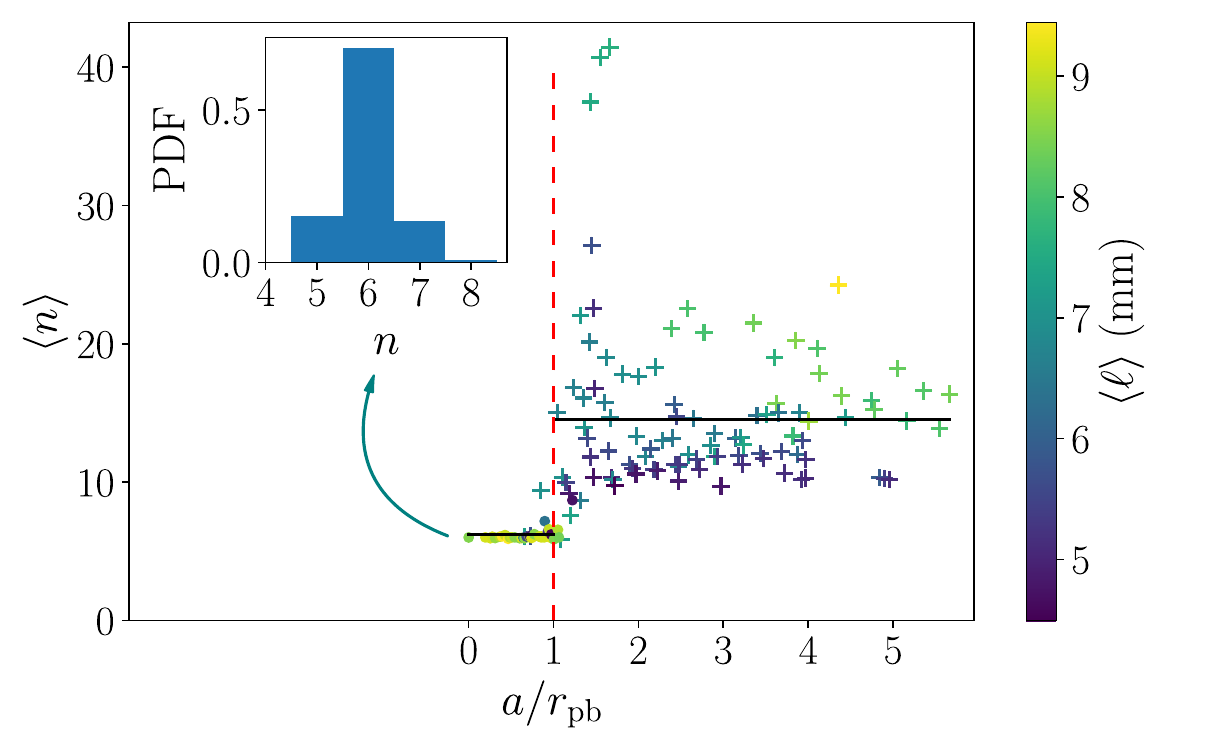}
    \caption{Average number of segments $\langle n \rangle $ as a function of the dimensionless size of asperities $a/r_{\rm pb}$.
    Horizontal lines are mean values for $a/r_{\rm pb}<1$ and $a/r_{\rm pb}>1$, and are equal to 6.2 and 14.6 respectively.
    The markers encode the visual classification made in figure~\ref{fig:visualization} with $\circ$ for straight PPBs and $+$ for the zigzag morphology.
    Images illustrating the PPBs in the zigzag regime are provided in SI.
    The inset is the PDF of the number of segments for footprints satisfying $a/r_{\rm pb} < 1$.
    }
    \label{fig:number}
\end{figure}

To represent the variation of the number of segment $n$, we start by analyzing the space of the control parameters, which is composed of the size of asperities $a$,  the bubble radius $R$, and the liquid fraction $\varphi$.
The two last parameters are related to the bulk properties of the foam.
The corresponding surface properties are the PPBs  perimeter and width.
As the range of liquid fractions in our experiment corresponds to dry foams, geometrical considerations show that the radius of curvature of the Plateau borders is $r_{\rm pb} = R \sqrt{\varphi/0.33}$, which is a good estimate of the width \cite{Cantat2013b}.
Since $\ell \gg \{a, r_{\rm pb} \}$, it is natural to construct a dimensionless microscopic lengthscale $a/r_{\rm pb}$.

In figure~\ref{fig:number}, we plot the number of segments $\langle n \rangle$ averaged over several footprints in the same conditions as a function of $a/r_{\rm pb}$ with the average perimeter $\langle\ell\rangle$ encoded by a color map.
We observe two regimes.
For $a/r_{\rm pb}<1$, the number of segments per bubble is typically $\langle n \rangle = 6\pm 1$ as illustrated in the inset of figure~\ref{fig:number}.
This corresponds to the widely observed morphology of bubbles in contact with smooth surfaces \cite{Matzke1946,Wang2009}.
Bubble footprints are mainly hexagonal, and some of them are pentagonal or heptagonal.
For $a/r_{\rm pb}>1$, footprints have more than 8 segments and can be up to 40 segments on average with a characteristic mean value of 14.6 segments.
Figure~\ref{fig:number} suggests no evident correlation between the number of segments and the microscopic dimensionless parameter  $a/r_{\rm pb}$.
However, we clearly observe an increasing trend with the macroscopic footprint perimeter $\langle \ell \rangle$, and thus on the bubble radius $R$ (Fig.~{\ref{fig:visualization}}(e)).


To understand the transition at $a/r_{\rm pb}\simeq 1$, we supplemented the fluorescence microscopy with visualizations by confocal microscopy (Leica, TCS SP8).
We used an optical adaptation allowing measurements in a direction perpendicular to gravity \cite{Mariot2018}, as for the visualization by fluorescence microscopy.
For imaging purposes, we also dissolved Nile red dye to the PDMS when preparing the surface.

\begin{figure}[ht]
    \centering
    \includegraphics[width=\linewidth]{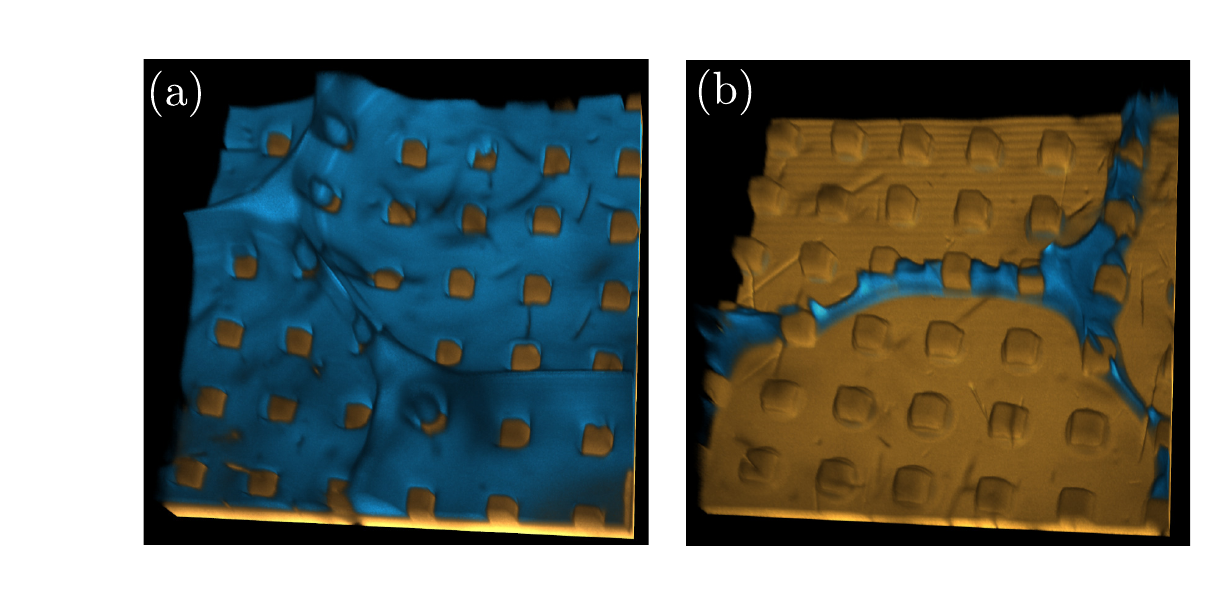}
    \includegraphics[width=\linewidth]{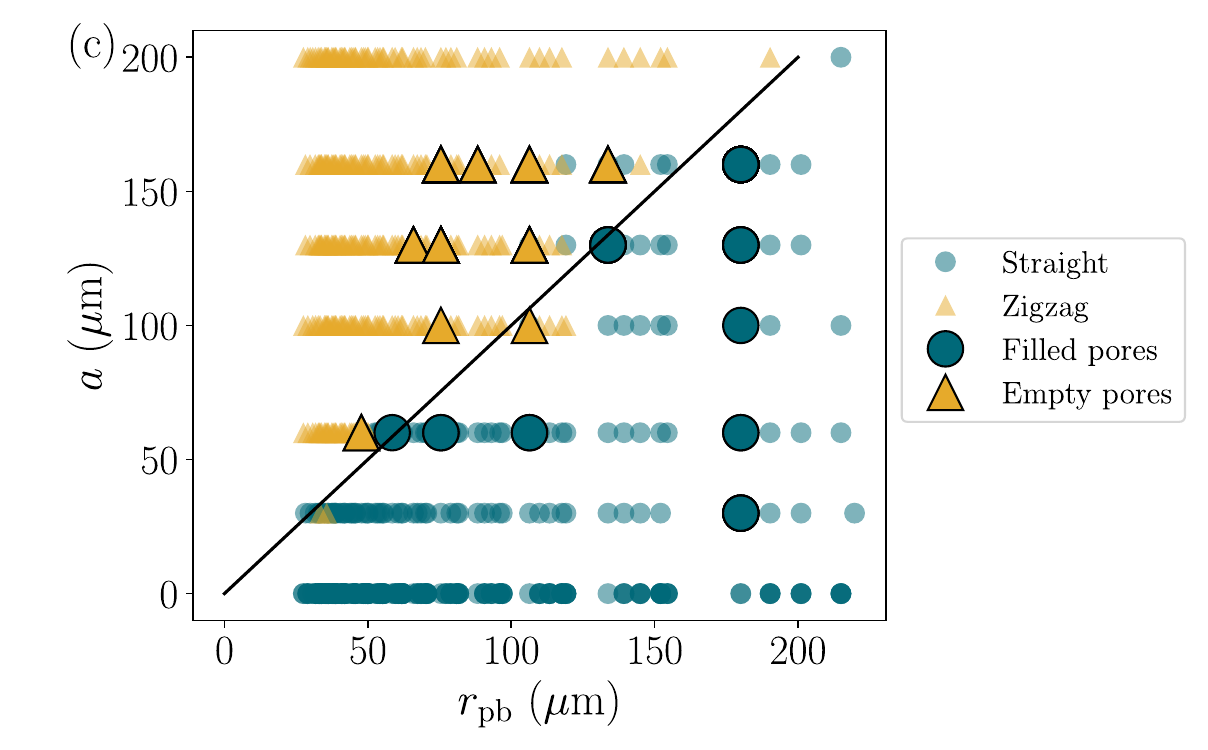}
    \caption{
    (a-b) Images obtained by confocal microscopy where the textured surface is represented in yellow and the liquid is in blue where $a=$ 100 $\mu$m.
    (a) for $r_{\rm pb}=$ 180 $\mu$m, the surface is filled by a liquid film of the thickness of the pillars, and (b) the surface pores are empty ($r_{\rm pb}=$ 75 $\mu$m).
    The width of the imaged surface is 1.2 mm.
    (c) Phase diagram of the PPB morphology representing $a$ versus $r_{\rm pb}$.
    Large blue circles are for surfaces covered by a liquid film, illustrated in (a) and large yellow triangles are for dry surfaces as shown in (b).
    The black line is the equality between axes.
    }
    \label{fig:confocal}
\end{figure}

In figure~\ref{fig:confocal}, we report our observations where two configurations are encountered depending on the geometric parameter $a$ and the radius of the Plateau borders $r_{\rm pb}$.
For $a < r_{\rm pb}$ (Fig.~\ref{fig:confocal}(a)), the surface is fully covered by a liquid film in addition to the PPBs whereas for $a > r_{\rm pb}$ (Fig.~\ref{fig:confocal}(b)), the PPBs in blue are directly in contact with the textured surface in yellow.
In figure~\ref{fig:confocal}(c), we present the observations combining the PPB morphology and filling state of the pores, each being well-classified by the value $a/r_{\rm pb}$.
This wetting transition can be explained by the capillary pressure of the foam as follows.

 The capillary pressure of the meniscus between four pillars  scales as $\gamma/a$ \cite{Xiao2010}.
 Balancing this pressure with the pressure in the liquid foam $\gamma/r_{\rm pb}$ leads to a critical value for the existence of such inter-pillar meniscus, which scales as $a/r_{\rm pb}$.
This criterion is in agreement with our observations in figure~\ref{fig:confocal}(c) where the prefactor is close to unity.
Thus, we now understand that for $a/r_{\rm pb}<1$, the liquid film on the surface prevents the distortion of the PPBs whereas for $a/r_{\rm pb}>1$, the PPBs are between the asperities causing a zigzag morphology. the PPBs are trapped between asperities of larger height and spacing that causes a zigzag morphology to comply with the square lattice.

\begin{figure}[t]
    \centering
    \includegraphics[width=\linewidth]{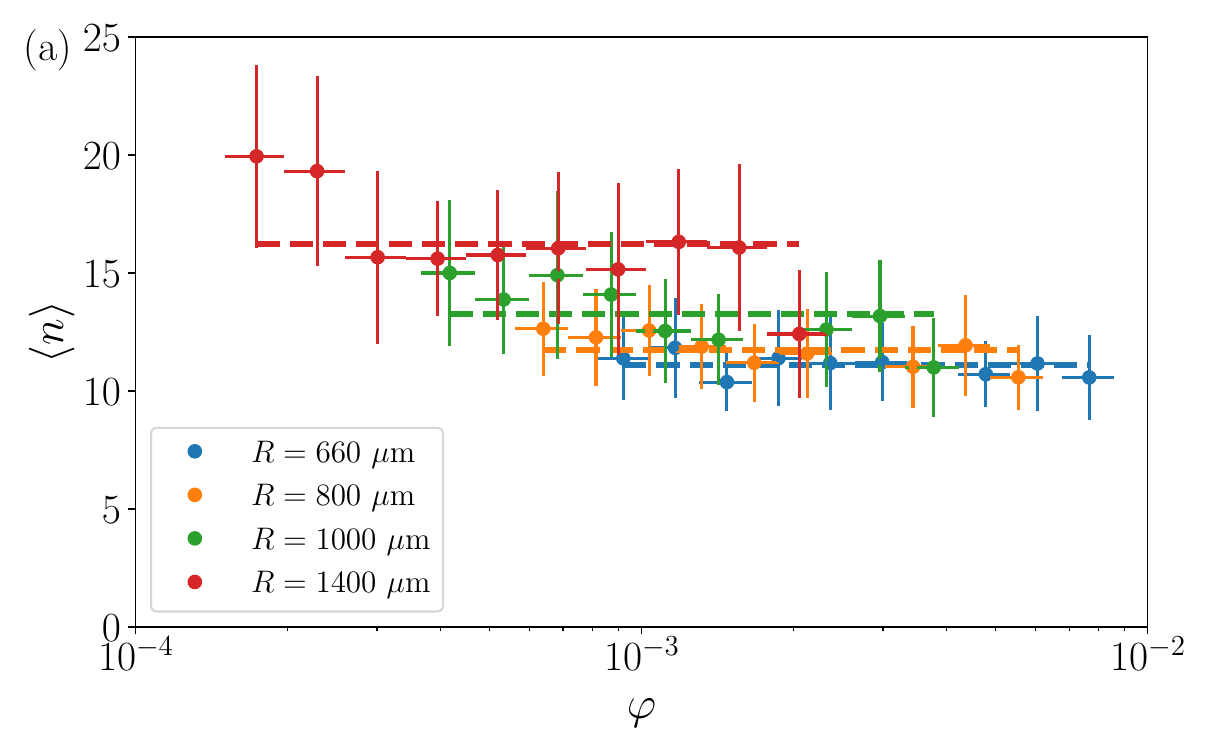}
    \includegraphics[width=\linewidth]{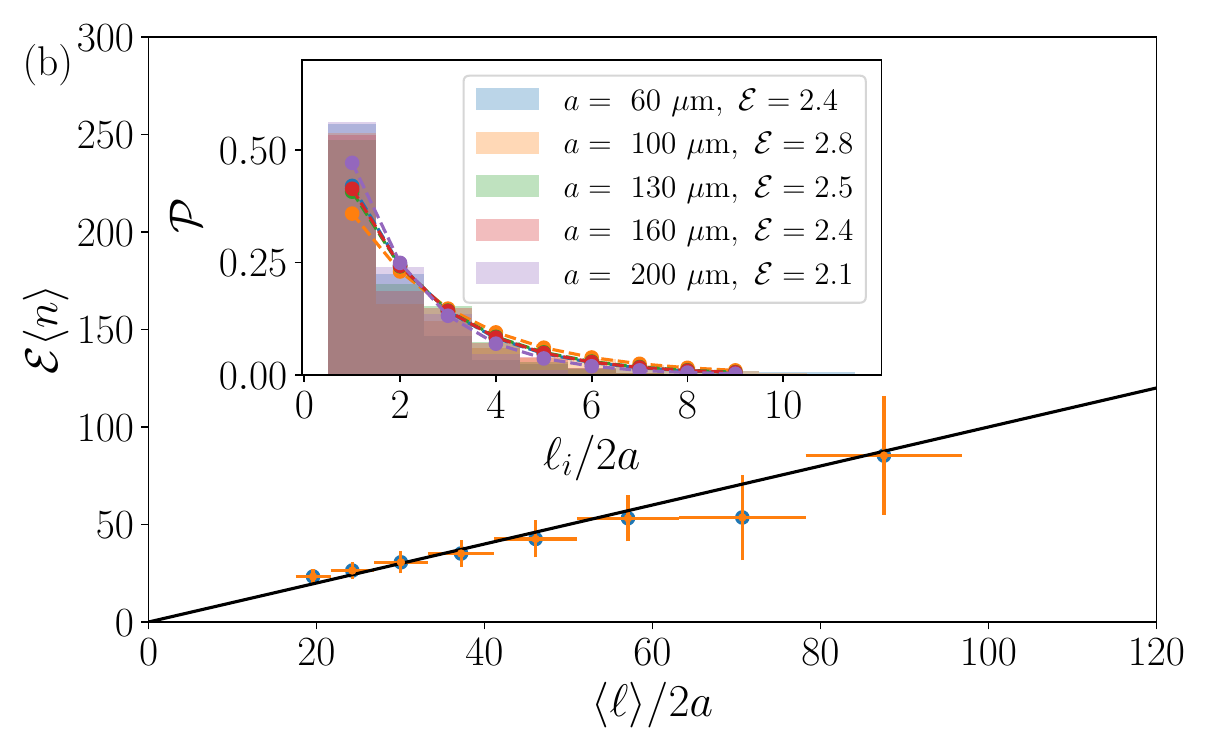}
    \caption{Statistics in the zigzag regime.   (a) The mean number of segments $\langle n \rangle$ is plotted against $\varphi$ for $a=130~\mu$m in the regime $a/r_{\rm pb} > 1$.
    The dashed lines are the mean values for each bubble radius $R$.
    Vertical error bars represent the standard deviation and horizontal ones the interval.
    (b) The main plot shows ${\cal E} \langle n \rangle$ as a function of the dimensionless parameter $\langle \ell \rangle / 2a$ for the entire dataset in the zigzag regime.
    The solid line represents equality between axes.
    The inset is the PDF ${\cal P}$ of the lengths $\ell_i/2a$ for five different sizes $a$.
    On both plots, the vertical error bar represents the standard deviation and the horizontal one is the range over which the data are regrouped.    }
    \label{fig:perimeter}
\end{figure}

As we noticed in figure~\ref{fig:number} that the number of footprint segments increases with the bubble perimeter in the zigzag regime, we propose to seek for a dimensionless number describing the zigzag regime.
In figure~\ref{fig:perimeter}(a), we observe that the number of segments $\langle n \rangle$ is independent of the liquid fraction $\varphi$ for different bubble radii in the regime $a/r_{\rm pb} > 1$.
Thus, we exclude an effect of the PPB size, which suggests that the number of segments depends only on the pattern parameters.

The inset of figure~\ref{fig:perimeter}(b) shows the probability distribution function (PDF) of the normalized  lengths of segments ${\tilde \ell}_i = \ell_i/2a$.
We choose the center to center distance of adjacent asperities as a normalization length.
The probability decreases with ${\tilde \ell}_i$ independently of the pattern size $a$.
From the PDFs, we compute the mean mathematical expectation ${\cal E}=\sum_i {\cal P}[{\tilde \ell}_i] {\tilde \ell}_i$, whose values are indicated in the inset of figure~\ref{fig:perimeter}(b) for each distribution.
The  mathematical expectation is not correlated  to the mesh size and is ${\cal E} = 2.4\pm0.2$.

To model the PDFs, we consider that the length of an edge $\ell_i$ is the result of Bernoulli trials.
From a segment of an edge, the next step over the mesh size $2a$ is made with a probability $p$ to get an aligned extension and $(1-p)$ to form a kink, independently of the previous orientation.
Thus, the probability mass function follows a geometric distribution \cite{Dekking2005}
\begin{equation}\label{eq:probability}
   {\cal P} = (1-p)^{{\tilde \ell}_i - 1} p,
\end{equation}
where by construction $p = 1/{\cal E}$.
Equation~\ref{eq:probability} is plotted over the distribution in the inset of figure~\ref{fig:perimeter}(b) with a good agreement.

The average length of an edge being $2a\, {\cal E}$, we deduce that the characteristic perimeter length  $\langle \ell \rangle$ is proportional to the mesh unit size $2a$, the number of footprint segments $\langle n \rangle$, and the  mathematical expectation ${\cal E}$, \textit{i.e.}
\begin{equation}\label{eq:master_equation}
    \langle \ell \rangle = 2a\, {\cal E} \langle n \rangle.
\end{equation}
From our experimental data, we successfully plot, in figure~\ref{fig:perimeter}(b), the dimensionless form of equation~\ref{eq:master_equation} without any fitting parameter.
Interestingly, equation~\ref{eq:master_equation} can serve as a prediction of the mean number of footprint segments.
Indeed, the mathematical expectation ${\cal E}$ is independent of size $a$ and the foam parameters.
As shown in figure~\ref{fig:visualization}(e), the mean perimeter $\langle \ell \rangle$ is proportional to the bubble radius $R$ such that there is a direct relation between $R$ and the average number of segments $\langle n\rangle$.

%
%
\section{Conclusion}

In conclusion, when a foam is placed on a surface with regularly spaced asperities, we observe a straight to zigzag transition of the PPBs.
We demonstrated that this transition depends on the dimensionless parameter $a/r_{\rm pb}$ that compares the Laplace pressure of the foam with the capillary pressure of the asperities.
Below unity, the filled pores offer a smooth-like surface to the foam, while above, the PPBs are between the asperities imposing a tortuosity.
In this latter regime, we rationalized that the number of  PPB segments increases linearly with $\langle \ell\rangle/2a$, with a prefactor that is the mathematical expectation.
Future studies will be necessary to expand these findings, exploring a wider range of geometrical parameters of the pillars such as the height, the spacing, the spatial organization as well as the shape.

So far, the motion of bubbles on surface decorated with pillars has been studied for bubbly liquid where the zigzag regime is not observed \cite{Germain2016}.
The motion of surface bubbles of a dry foam focused on surfaces randomly covered by glued beads, for which sliding, stick-slip and anchored behaviors were observed \cite{Marchand2020}.
In forthcoming studies, we plan to analyze how regularly spaced asperities can affect the dynamics of bubbles on the solid surface depending on the PPB morphology.
Additionally, since the coarsening dynamics is primarily affected by the topology, we anticipate that such observations have an impact on the aging process of foams.
Eventually, a heterogeneous coarsening, with a different dynamics close to rough surfaces could lead to structural gradients in foams.

\paragraph{Acknowledgments}
We kindly thank Antoine Boury and Raphael Weil for their guidance on the soft photolithography.
We also thank the SoMaC $\&$ CoMic platform for the access to the confocal microscope.
We acknowledge for funding support from the French Agence Nationale de la Recherche in the framework of project AsperFoam - 19-CE30-0002-01.


\begin{thebibliography}{10}

\bibitem{Plateau1873}
J.~A.~F. Plateau.
\newblock {\em Statique exp{\'e}rimentale et th{\'e}orique des liquides soumis
  aux seules forces mol{\'e}culaires}, volume~2.
\newblock Gauthier-Villars, 1873.

\bibitem{Taylor1976}
J.E. Taylor.
\newblock The structure of singularities in soap-bubble-like and soap-film-like
  minimal surfaces.
\newblock {\em Ann. Math}, 103:489, 1976.

\bibitem{Weaire1999}
D.~L. Weaire and S.~Hutzler.
\newblock {\em The physics of foams}.
\newblock Oxford University Press, 1999.

\bibitem{Cantat2013b}
I.~Cantat, S.~Cohen-Addad, F.~Elias, F.~Graner, R.~H{\"o}hler, O.~Pitois,
  F.~Rouyer, A.~Saint-Jalmes, R.~Flatman, and S.~Cox.
\newblock {\em Foams: Structure and Dynamics}.
\newblock OUP Oxford, 2013.

\bibitem{Matzke1946}
E.~B Matzke.
\newblock The three-dimensional shape of bubbles in foam-an analysis of the
  role of surface forces in three-dimensional cell shape determination.
\newblock {\em American journal of Botany}, pages 58--80, 1946.

\bibitem{Jouanlanne2022}
M.~Jouanlanne, A.~Egelé, D.~Favier, W.~Drenckhan, J.~Farago, and
  A.~Hourlier-Fargette.
\newblock Elastocapillary deformation of thin elastic ribbons in 2d foam
  columns.
\newblock {\em Soft Matter}, 18:2325--2331, 2022.

\bibitem{Guidolin2023}
C.~Guidolin, J.~Mac~Intyre, E.~Rio, A.~Puisto, and A.~Salonen.
\newblock Viscoelastic coarsening of quasi-2d foam.
\newblock {\em Nature Communications}, 14(1):1125, 2023.

\bibitem{Glazier1987}
J.~A. Glazier, S.~P. Gross, and J.~Stavans.
\newblock Dynamics of two-dimensional soap froths.
\newblock {\em Physical Review A}, 36(1):306, 1987.

\bibitem{Roth2013}
A.E. Roth, C.D. Jones, and D.J. Durian.
\newblock Bubble statistics and coarsening dynamics for quasi-two-dimensional
  foams with increasing liquid content.
\newblock {\em Physical Review E}, 87(4):042304, 2013.

\bibitem{Cubaud2001}
T.~Cubaud and M.~Fermigier.
\newblock Faceted drops on heterogeneous surfaces.
\newblock {\em Europhysics Letters}, 55(2):239, jul 2001.

\bibitem{Bico2001}
J.~Bico, C.~Tordeux, and D.~Quéré.
\newblock Rough wetting.
\newblock {\em Europhysics Letters}, 55(2):214, jul 2001.

\bibitem{Courbin2007}
L.~Courbin, E.~Denieul, E.~Dressaire, M.~Roper, A.~Ajdari, and H.~A. Stone.
\newblock Imbibition by polygonal spreading on microdecorated surfaces.
\newblock {\em Nature Materials}, 6(9):661--664, 2007.

\bibitem{Kim2011b}
S.~J. Kim, M.-W. Moon, K.-R. Lee, D.-Y. Lee, Y.~S. Chang, and H.-Y. Kim.
\newblock Liquid spreading on superhydrophilic micropillar arrays.
\newblock {\em Journal of Fluid Mechanics}, 680:477–487, 2011.

\bibitem{Gauthier2014}
A.~Gauthier, M.~Rivetti, J.~Teisseire, and E.~Barthel.
\newblock Finite size effects on textured surfaces: Recovering contact angles
  from vagarious drop edges.
\newblock {\em Langmuir}, 30(6):1544--1549, February 2014.

\bibitem{Kim2016}
J.~Kim, M.-W. Moon, and H.-Y. Kim.
\newblock Dynamics of hemiwicking.
\newblock {\em Journal of Fluid Mechanics}, 800:57–71, 2016.

\bibitem{DeGennes1985}
P.-G. De~Gennes.
\newblock Wetting: statics and dynamics.
\newblock {\em Rev. Mod. Phys.}, 57(3):827--863, Jul 1985.

\bibitem{Raphael1989}
E.~Raphaël and P.-G. de~Gennes.
\newblock {Dynamics of wetting with nonideal surfaces. The single defect
  problem}.
\newblock {\em The Journal of Chemical Physics}, 90(12):7577--7584, 06 1989.

\bibitem{Quere2008a}
D.~Quéré.
\newblock Wetting and roughness.
\newblock {\em Annu. Rev. Mater. Res.}, 38(1):71--99, 2008.

\bibitem{Gauthier2013}
A.~Gauthier, M.~Rivetti, J.~Teisseire, and E.~Barthel.
\newblock Role of kinks in the dynamics of contact lines receding on
  superhydrophobic surfaces.
\newblock {\em Phys. Rev. Lett.}, 110:046101, Jan 2013.

\bibitem{Mannetje2014}
D.~Mannetje, S.~Ghosh, R.~Lagraauw, S.~Otten, A.~Pit, C.~Berendsen, J.~Zeegers,
  D.~van~den Ende, and F.~Mugele.
\newblock Trapping of drops by wetting defects.
\newblock {\em Nature Communications}, 5(1):3559, 2014.

\bibitem{Johnson2022}
P.~Johnson, V.~Starov, and A.~Trybala.
\newblock Foam flow through porous media.
\newblock {\em Current Opinion in Colloid \& Interface Science}, 58:101555,
  2022.

\bibitem{Wang2009}
Y.~Wang and S.~J. Neethling.
\newblock The relationship between the surface and internal structure of dry
  foam.
\newblock {\em Colloids and Surfaces A: Physicochemical and Engineering
  Aspects}, 339(1):73--81, 2009.

\bibitem{Xiao2010}
R.~Xiao, R.~Enright, and E.~N. Wang.
\newblock Prediction and optimization of liquid propagation in micropillar
  arrays.
\newblock {\em Langmuir}, 26(19):15070--15075, 2010.

\bibitem{Dekking2005}
F.~M. Dekking, C.~Kraaikamp, H.~P. Lopuha{\"a}, and L.~E. Meester.
\newblock {\em A Modern Introduction to Probability and Statistics:
  Understanding why and how}, volume 488.
\newblock Springer, 2005.

\bibitem{Germain2016}
D.~Germain and M.~Le Merrer.
\newblock Bubbles slipping along a crenelated wall.
\newblock {\em EPL}, 115(6):64005, 2016.

\bibitem{Marchand2020}
M.~Marchand, F.~Restagno, E.~Rio, and F.~Boulogne.
\newblock Roughness-induced friction in liquid foams.
\newblock {\em Physical Review Letters}, 124:118003, 2020.

\bibitem{Boulogne2023}
F.~Boulogne, E.~Rio, and F.~Restagno.
\newblock Evaporation-induced temperature gradient in a foam column.
\newblock {\em Langmuir}, 39(40):14256--14262, 2023.

\bibitem{Mariot2018}
S.~Mariot, M.~Schneider, A.~Salonen, and W.~Drenckhan.
\newblock Optical adaptation of confocal microscopes for arbitrary imaging
  angles—and its application to sedimentation/creaming in dispersions.
\newblock {\em Measurement Science and Technology}, 29(12):127001, oct 2018.

\end{thebibliography}

\bibliographystyle{unsrt}

\newpage\clearpage


\section*{Supplementary material}

# Supplementary information:
# Straight to zigzag transition of foam pseudo Plateau borders on textured surfaces

Alexis Commereuc[1], Sandrine Mariot[1], Emmanuelle Rio[1], and François Boulogne[1]

[1]Université Paris-Saclay, CNRS, Laboratoire de Physique des Solides, 91405, Orsay, France.

## 1 High resolution images from Figure 1

We provide high resolution images of Figure 1 to show the details of the zigzag pattern and the counting of the segment number. Figure S1 corresponds to Figure 1(a) and Figure S2 to Figure 1(b).

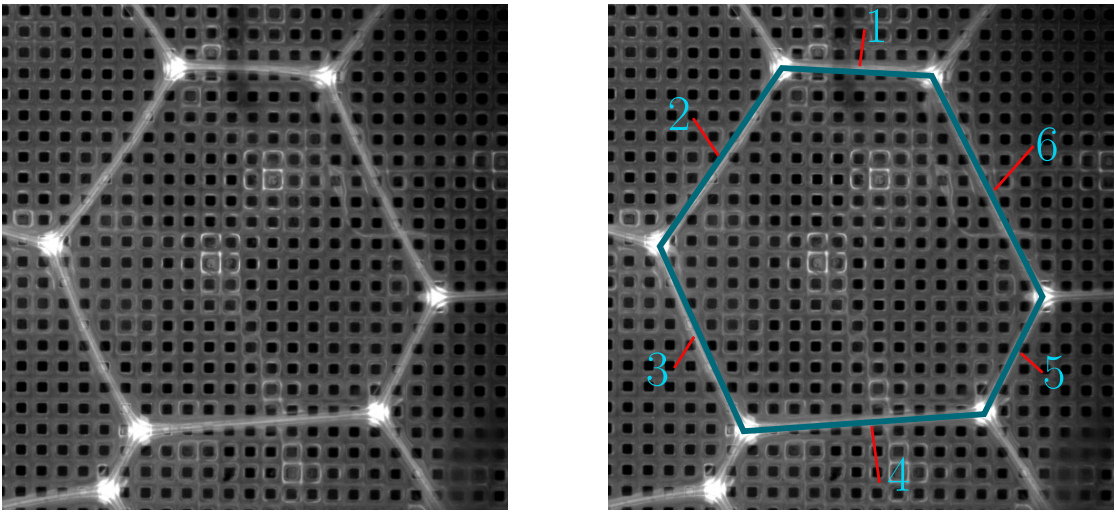

Figure S1: High resolution version of Figure 1(a) where the image on the left is the original image and on the right, we superimposed the segments. Here, $n = 6$, for a bubble radius $R = 1000$ $\mu$m, a surface parameter $a = 60$ $\mu$m and a liquid fraction $\varphi = 0.28$ %, $i.e.$ $a/r_{\rm pb} = 0.65$.

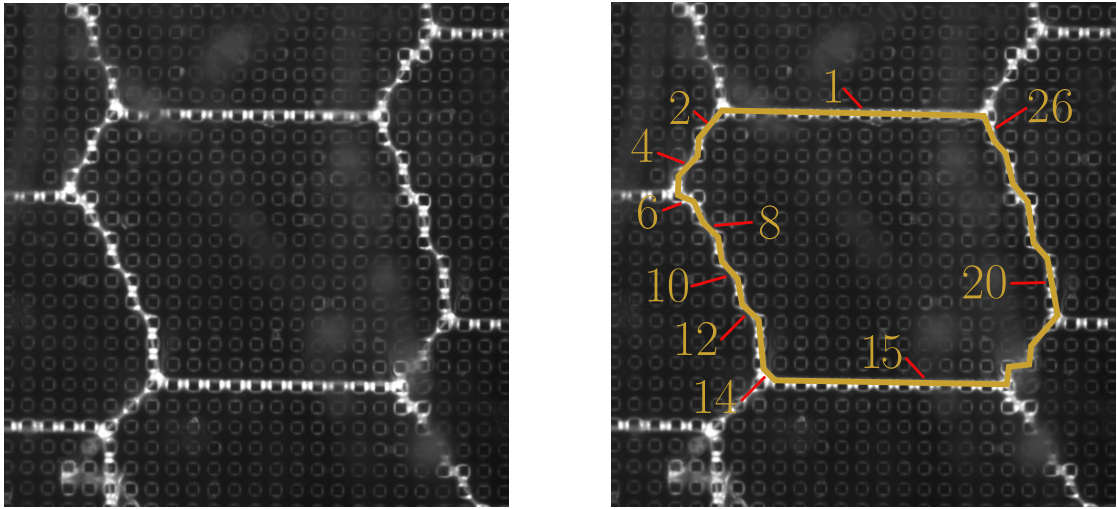

Figure S2: High resolution version of Figure 1(b) where the image on the left is the original image and on the right, we superimposed the segments. Here, $n = 26$, for a bubble radius $R = 1000$ $\mu$m, a surface parameter $a = 60$ $\mu$m and a liquid fraction $\varphi = 0.03$ %, i.e. $a/r_{\rm pb} = 2.0$.

## 2 High resolution images for $a/r_{\rm pb} > 1$

To illustrate the variety of zigzag patterns we observed, and more specifically the few cases where a large number of segments is observed in figure 2, we provide examples in figure S3 of the zigzag patterns. The purpose of these images is also to explain the choices we made to count the number of segments.

In figure S3(a,b), we show examples of bubbles for respectively $a/r_{\rm pb} = 1.61$ and $a/r_{\rm pb} = 2.84$. We observe that the PPB can bridge either direct neighboring pillars or diagonally. In both cases, the number of segments remains typically lower than 25.

However, in figure S3(c), we show an example where we detected a large number of segments ($n = 35$). The magnified regions illustrate the difficulty to define the segments at the transition. Here, we observe mostly *staircase* patterns generating a large number of segments. This is characteristic of the highest number of edges (typically $n > 25$) encountered solely just above the straight to zigzag transition.

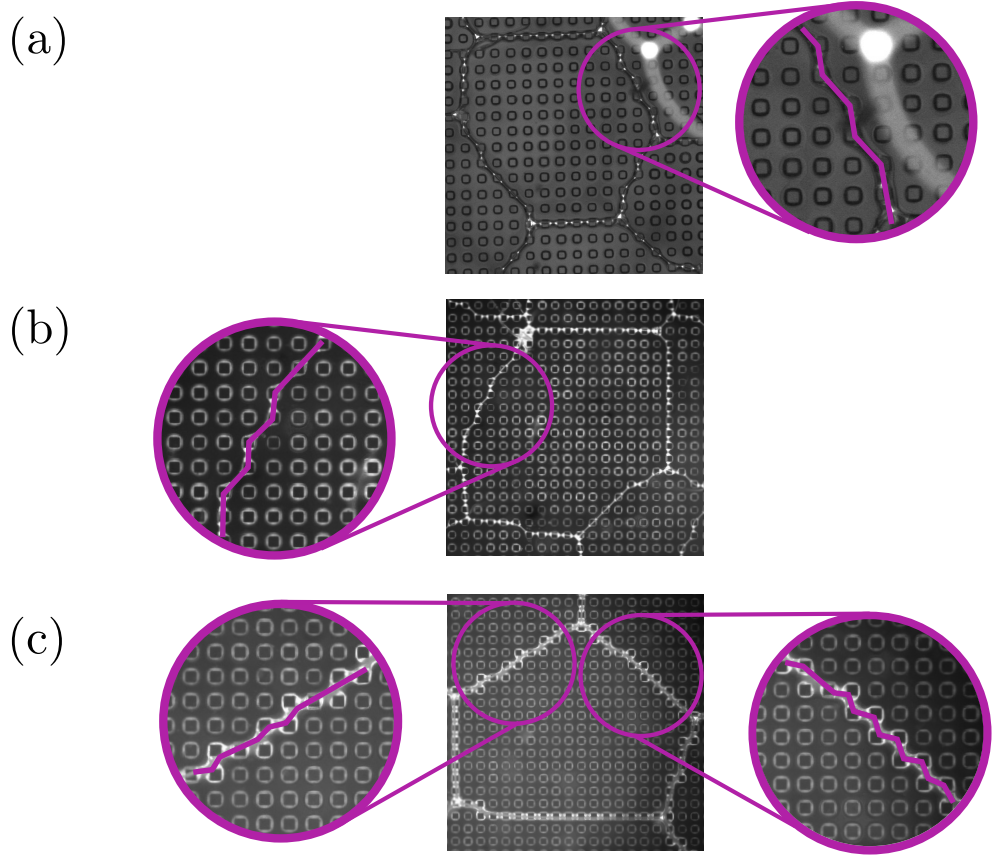

Figure S3: Photographs of two zigzag patterns near and long after the transition. (a) the parameters are $a = 100$ $\mu$m, $R = 1400$ $\mu$m and $\varphi_l = 0.06$ %, *i.e.* $a/r_{pb} = 1.61$. We obtained $n = 21$. (b) the parameters are $a = 130$ $\mu$m, $R = 1400$ $\mu$m and $\varphi_l = 0.03$ %, *i.e.* $a/r_{pb} = 2.84$. We obtained $n = 19$. (c) The parameters are $a = 130$ $\mu$m, $R = 1400$ $\mu$m and $\varphi_l = 0.09$ %, *i.e.* $a/r_{pb} = 1.59$. We obtained $n = 35$.



\end{document}